\documentclass[fleqn,twoside]{article}
\usepackage{espcrc2}
\usepackage{amssymb}

\usepackage{graphicx}
\usepackage{epsfig}
\usepackage[figuresright]{rotating}

\newcommand{\AmS}{{\protect\the\textfont2
  A\kern-.1667em\lower.5ex\hbox{M}\kern-.125emS}}
\newcommand{\Vus}{\ensuremath{|V_{us}|}}
\newcommand{\dEM}[1]{\ensuremath{\delta_{\rm EM}^{#1}}}
\newcommand{\dSU}[1]{\ensuremath{\delta_{\rm SU(2)}^{#1}}}

\title{Dispersive representation of the scalar and vector K$\pi$ form factors for $\tau \rightarrow  K \pi \nu_\tau$ and $K_{\ell 3}$ decays}

\author{V. Bernard\address{Groupe de Physique Th\'eorique, IPN,
Universit\'e de Paris Sud-XI/CNRS, F-91406 Orsay, France},
D. R. Boito\address{Grup de F\'isica Te\`orica and IFAE, Universitat Aut\`onoma de Barcelona, E-08193 Bellaterra (Barcelona), Spain} 
and
E. Passemar\thanks{Speaker}\address{
IFIC, Universitat de Val\`encia - CSIC,
Apartat de Correus 22085, E-46071 Val\`encia, Spain}}
       
\begin{document}

\begin{abstract}
Recently, the $\tau \rightarrow  K \pi \nu_\tau$ decay spectrum has been measured by
the Belle and BaBar collaborations. In this work, we present an
analysis of such decays introducing a dispersive parametrization for the
vector and scalar K$\pi$ form factors. This allows for precise
tests of the Standard Model. For instance, the determination of
$f_+(0)|V_{us}|$ from these decays is discussed. 
A comparison and a combination of these results with the analyses of the $K_{\ell3}$ decays
is also considered.
\vspace{1pc}
\end{abstract}

\maketitle

\section{INTRODUCTION}

Despite the great success of the Standard Model (SM), there are indications that 
it is the effective theory of a more fundamental theory with new degrees of freedom appearing at the TeV scale. 
There exist two main  approaches to look for physics beyond the Standard Model:
direct searches for new particles (Charged Higgs, Supersymmetric particles, Z', W'...) at high ener-gy
colliders and indirect searches, for instance in flavour physics, through precision experiments. 
 
We will follow here the second approach and test the SM studying the $\tau \rightarrow  K \pi \nu_\tau$
and $K_{\ell3}$ decays.
For that a very precise knowledge of the K$\pi$ form factors is necessary. 
Until recently, 
experimental information on these form factors was only coming from 
the $K_{\ell3}$ ($K \rightarrow \pi \ell \nu_\ell$, $\ell=\mu, e$) decay measurements~\cite{Flavia}. 
But new high statistic measurements of 
the $\tau \rightarrow  K \pi \nu_\tau$ decays from Belle~\cite{Belle} and BaBar~\cite{BaBar} 
make possible to constrain them further as the 
relevant hadronic matrix element in this decay  
corresponds to the crossed channel with respect to the $K_{\ell 3}$, 
\begin{eqnarray} 
\langle  K \pi | \bar{s}\gamma_{\mu}u | 0 \rangle = -\frac{\Delta_{K\pi}}{s} (p_K+p_\pi)_\mu f_0^{K \pi } (s) + \nonumber \\
\left[(p_K-p_\pi)_\mu + \frac{\Delta_{K\pi}}{s}(p_K+p_\pi)_\mu \right] f^{K \pi}_+ (s)~.
\end{eqnarray}
$s=(p_K+p_\pi)^2=(p_\tau-p_\nu)^2$ is the exchanged four-momentum and $\Delta_{K\pi}= m_K^2-m_\pi^2$. 
The vector form factor $f_+(s)$ represents the $P$-wave projection of $\langle 0 | \bar{s}\gamma_{\mu}u | K \pi \rangle$ 
whereas the scalar form factor $f_0(s)$ 
describes the $S$-wave projection, and 
one has 
$f_0(0)=f_+(0)$. 
These measurements motivated several 
analyses~\cite{Jamin,Bachir1,Diogo1,Diogo2} 
introducing some representations for the shape of the vector form factor 
$\bar f_+$, 
\begin{equation}
\bar f_{+,0}(s) = \frac{f_{+,0}(s)}{f_{+,0}(0)}~,
\label{eq:Normff}
\end{equation}
relying on fundamental properties such as ana-lyticity, unitarity and 
short distance QCD. 
In Ref.~\cite{Diogo2}, a combined analysis of $\tau \rightarrow K\pi \nu_\tau$ and $K_{\ell 3}$ decays has also 
been performed.  
In all these studies  $\bar f_0(s)$ has been taken from some
models. 
In this work, we investigate the constraints on the K$\pi$ form factors coming from $\tau \rightarrow K\pi \nu_\tau$ 
and $K_{\ell 3}$ decays 
using a dispersive representation for both $\bar f_0(s)$~\cite{BOPS06,BOPS09} and $\bar f_+(s)$.  
Following Refs.~\cite{Diogo1,Diogo2}, we use 
three times subtracted dispersive relations, howe-ver in comparison to these references 
we will impose the short distance constraints from perturbative QCD.  
Furthermore and more importantly, we extract $f_+(0)$\Vus~from the $\tau \rightarrow K\pi \nu_\tau$ 
decay measurements; it was an input in the previous analyses.
\section{TESTS OF THE STANDARD MODEL}
The knowledge of the K$\pi$ form factors allows for precision
tests of the SM. 
\subsection{Extraction of \Vus}  
The CKM mixing matrix element \Vus~has been very precisely 
determined from $K_{\ell 3}$ decays~\cite{Flavia}. 
However, it is also possible to extract it from the measurement of the 
$\tau \rightarrow K\pi \nu_\tau$ decays. 
Indeed, the $\tau \rightarrow K\pi \nu_\tau$ and $K_{\ell 3}$ decay rates can be expressed as 
\begin{eqnarray}
\Gamma_{i} & \!\!=& \!\! \ G_F^2 \mathcal{N}_i \, C_{Ki}^2 S_{\rm EWi} 
\left(\Vus f_+^{K^0 \pi^-}(0) \right)^2 I_{K}^i \nonumber \\
& & \times \left(1 + \dEM{i} + \dSU{i} \right)^2,
\label{eq:M}
\end{eqnarray}
with $i$ standing for $K_{\ell 3}$ or $\tau \rightarrow K\pi \nu_\tau$. 
The expression of the quantities entering Eq.~(\ref{eq:M}) for $K_{\ell 3}$ decays 
can be found in Ref.~\cite{Flavia}. We only give the ones for 
$\tau \rightarrow K\pi \nu_\tau$ below.  
$\mathcal{N}_i$ is a normalization 
coefficient ($\mathcal{N}_\tau = m_\tau^3/(48 \pi^3)$), $G_F$ the Fermi constant and  $C_{Ki}$ a Clebsch-Gordan coefficient 
($C_{K, \tau} =$ $1/\sqrt{2}$ for $K^0$ and $1/2$ for $K^-$).
A very precise determination of \Vus~requires: 

i) a very accurate measurement of  $\Gamma_{i}$, 

ii) a very precise calculation of the phase space integrals $I_{K}^i$ that probe the energy dependence of the form factors 
\begin{eqnarray}
\label{eq:ITau}
I_{K}^\tau & \!\!= &\!\! \int_{s_{K\pi}}^{m_\tau^2}\!\! \frac{ds}{s\sqrt{s}}\, 
\left(1-\frac{s}{m^2_\tau}\right)^2 
\left[\left(1+\frac{2s}{m^2_\tau}\right) \right . \\
& & \left. \times q^3_{K\pi}(s)\bar f^2_+(s)+ 
\frac{3 \Delta_{K\pi}^2}{4s}q_{K\pi}(s)\,\bar f_0^2(s)\right], \nonumber 
\end{eqnarray}
with $s_{K\pi}=(m_K+m_\pi)^2$ and $q_{K\pi}$ the kaon momentum in the rest frame of the hadronic system
\begin{equation}
q_{K\pi}=\frac{1}{2\sqrt{s}} \sqrt{\left(s- s_{K\pi}\right) \left(s-t_{K\pi} \right)}\times \theta \left(s- s_{K\pi}\right),
\end{equation}
with $t_{K\pi}=(m_K-m_\pi)^2$.

iii) a good knowledge of
the radiative corrections: the electroweak short-distance $\left( S_{EWi} \right)$, 
electromagnetic long-distance $\left(\dEM{i}\right)$ and isospin breaking $\left(\dSU{i} \right)$ corrections. 
The radiative corrections have been precisely evaluated for the $K_{\ell 3}$ decays \cite{Kastner08,Cirigliano08}, but in the case of 
$\tau \rightarrow K \pi \nu_\tau$ only 
$S_{EW, \tau}= 1.0201$ \cite{Erler02} is known, $\dEM{\tau}$ and $\dSU{\tau}$ have not been computed yet. 
They are estimated to be of $\sim 1\%$~\cite{Francisco}. 

iv) a determination of the value of the form factor at zero momentum transfer $f_+(0)$. 
This value can be obtained either from Chiral Perturbation Theory (ChPT)~\cite{f+0ChPT,Kastner08} or from lattice calculations~\cite{Juettner}. 

\subsection{Callan-Treiman theorem}
Another interesting test of the Standard Model is provided by the low energy theorem from Callan and Treiman 
(CT)~\cite{Callan}. This theorem predicts the value of the scalar form factor at the so-called CT point, 
$s_\mathrm{CT} \equiv \Delta_{K \pi}$, 
\begin{equation}
C \equiv \bar f_0(\Delta_{K\pi})=\frac{f_K}{f_\pi}\frac{1}{f_+(0)}+ \Delta_{CT}~,
\label{eq:CTrel}
\end{equation}
where $f_{K, \pi}$ are the kaon and pion decay constants respectively. 
$\Delta_{CT} \sim  {\mathcal{O}} (m_{u,d}/4 \pi F_{\pi})$ is a small correction computed in the 
framework of chiral perturbation theory~\cite{deltaCTChPT,Kastner08}. 
The test consists in determining the quantity $r$:
\begin{equation}
r =(C-\Delta_{CT}) \times \left. \left( \frac{f_\pi \cdot f_+(0)}{f_K}\right) \right|_{SM}~,
\end{equation}
where $f_K/(f_\pi \cdot f_+(0))|_{SM}$ is obtained from the branching fractions 
$\Gamma_{K^+_{\mu \nu}}/\Gamma_{\pi^+_{\mu \nu}}$ and the $\Gamma_{K_{Le3}}$ 
measurements assuming the standard electroweak couplings (CKM) 
while the value of $C$ is directly extracted from $\tau$ or 
$K_{\mu3}$ decay analyses\footnote{The scalar form factor is only measurable from the $K \rightarrow \pi \mu \nu_\mu$ being suppressed 
in the Dalitz plot density formula by $m_\ell^2/m_K^2$.}. 
A value of $r$  different from unity would indicate the presence of physics beyond the SM 
such as for instance right-handed quark currents~\cite{BOPS06} or 
a charged Higgs~\cite{Flaviaold}. For a determination of $r$ from $K_{\ell3}$
decays, see Refs.~\cite{Flaviaold,PassemarKaon09}. 

\section{\mbox{DISPERSIVE~REPRESENTATION~OF} THE K$\pi$ FORM FACTORS}
To determine $\bar f_+(s)$ and $\bar f_0(s)$, 
fits to the measured $K_{\ell3}$ or $\tau$ decay distributions are performed assuming a parametrization for the form factors. 
Until recently, for the $K_{\ell3}$ decays, the experimental collaborations were using a parametrization relying on a 
Taylor expansion 
\begin{equation}
\bar f_{+,0}^{Tayl}(s) = 1 + \lambda'_{+,0} \frac{s}{m_\pi^2} + \frac{1}{2}\lambda''_{+,0} \left(\frac{s}{m_\pi^2}\right)^2 + \ldots~,
\label{eq:Taylor}
\end{equation}
where $\lambda'_{+,0}$ and $\lambda''_{+,0}$ are the slope and curvature of the form factors respectively, or a pole parametrization. 
For $\tau$ decays ($s_{K\pi} \equiv (m_K + m_\pi)^2 < s < m_\tau^2$), 
the experimental analyses rely on a parametrization involving a sum of Breit-Wigner functions. 
While the use of such a parametrization, assuming the dominance of resonances for the vector form factor, 
is in good agreement with the data, for the scalar form factor there is no clear 
dominance of single resonances.  

Following previous work~\cite{Jamin,Bachir1,Diogo1,Diogo2,BOPS06,BOPS09}, we will use dispersive
relations which will allow us to describe simultaneously the physical
region of $K_{\ell 3}$ and $\tau \rightarrow K \pi \nu_\tau$ decays. 
\subsection{Vector form factor}
Following Ref.~\cite{Diogo1}, we write a dispersion relation for ln$\bar f_+(s)$ with three subtractions at $s=0$ leading 
to\footnote{$\bar f_+(s)$ is assumed not to have any zero.}
\begin{eqnarray}
\bar f_+(s) &\!\! = \!\! & \exp \left[ \lambda_+' \frac{s}{m_\pi^2}+\frac{1}{2} \left( \lambda_+'' - \lambda_+'^2 \right) 
\left( \frac{s}{m_\pi^2} \right)^2 \right. \nonumber \\
& & \left . +  \frac{s^3}{\pi} \int_{s_{K\pi}}^{\infty} \frac{ds'}{s'^3} \frac{\phi_+(s')}{(s'-s-i\epsilon)}\right]~.
\label{eq:Dispfv}
\end{eqnarray}
Use has been made of $\bar f_+(0)=1$ to fix one subtraction constant. 
$\lambda_+'$ and $\lambda_+''$ are the two other subtractions constants corresponding to 
the slope and curvature of the form factor, see Eq.~(\ref{eq:Taylor}). 
They are not known and are determined from a 
fit to the data. $\phi_+(s)$ represents the phase of $\bar f_+(s)$. 
According to Watson's theorem~\cite{Watson}, 
in the elastic region (here the inelasticity sets in with the opening of the 
first inelastic channel $K^*(892) \pi$), 
it is equal to the $P$-wave $I=1/2$ K$\pi$ scattering phase. 
Furthermore, $\bar f_{+}(s)$ vanishes as $\mathcal{O}(1/s)$ for large $s$~\cite{Lepage}, 
implying that $\phi_+(s) \stackrel{s\mapsto\infty}{\longmapsto } \pi$.  

In the $\tau$ decay region two resonances dominate, $K^*(892)$ and $K^*(1414)$.
As proposed in Refs.~\cite{Diogo1,Jamin}, one can use a parametrization 
for the vector form factor including the two resonances $K^*$ and $K^{*'}$\footnote{$K^*(1414)$ 
is denoted as $K^{*'}$ in the following.} to determine  $\phi_+(s)$: 
\begin{eqnarray}
\tilde f_+(s) & \!\! = & \! \! \frac{\tilde m_{K^*}^{2}-\kappa_{K^*} \tilde H_{K\pi}(0)+\beta s}{D(\tilde m_{K^*}, \tilde \Gamma_{K^*})}  \nonumber \\
& & - \frac{\beta s }{D(\tilde m_{K^{*'}}, \tilde \Gamma_{K^{*'}})}~,
\label{eq:Fphase}
\end{eqnarray}
with
\begin{eqnarray}
\hspace{-0.1cm}
D(\tilde m_{R}, \tilde \Gamma_{R})&\!\!\!\!\!=&\!\!\!\!\! \tilde m_R^2-s-\kappa_R~{\rm{Re}}~\tilde H_{K\pi}(s)-i\tilde m_R\tilde\Gamma_{R}(s). \nonumber 
\end{eqnarray} 
In this equation, 
$\tilde m_R$ and $\tilde \Gamma_R$ are some model parameters and the running width $\tilde \Gamma_{R}(s)$ is given by:
\begin{equation}
\tilde \Gamma_{R}(s)= \tilde \Gamma_R \frac{s}{\tilde m_R^2} \frac{\sigma_{K\pi}^3(s)}{\sigma_{K\pi}^3(\tilde m_R^2)}~,
\end{equation}
with $\sigma_{K\pi}(s)=2q_{K\pi}(s)/\sqrt{s}$. 
$\kappa_R$ is a parameter proportional to ${\rm{Im}}~\tilde H_{K \pi}(s)$
and $\tilde H_{K\pi}(s)$ corresponds to a well known $K\pi$ loop function in ChPT. 
$\beta$ is the mixing parameter between the two resonances. 
The mass $m_R$ and width $\Gamma_R$
of the two resonances are extracted from the complex pole position $s_R$
\begin{equation}
D(\tilde m_R, \tilde \Gamma_R)=0~~{\rm{for}}~~\sqrt{s_R}= m_R - \frac{i}{2} \Gamma_R~.
\end{equation}
One can take advantage of the $\tau \rightarrow K \pi \nu_\tau$ data for which 
the vector contribution dominates 
to determine the mass and width of the resonances from a fit to the data. As shown in Refs.~\cite{Jamin,Diogo1,Diogo2}, 
this leads to stringent constraints on the mass and width of the $K^*(892)$.
Note that the parametrization, Eq.~(\ref{eq:Fphase}) fulfills the short distance QCD 
properties and takes into account the $K\pi$ rescattering effects through the $\tilde H_{K\pi}$ terms, 
see Refs.~\cite{Jamin,Diogo1,Diogo2} for more details. 
Another remark concerns the $K^{*'}$. It predominantly decays in $K^*(892)\pi$~\cite{PDG} 
and work is in progress~\cite{BDP} to take into account 
this channel in the parametrization Eq.~(\ref{eq:Fphase}) following the coupled channel analysis 
performed in Ref.~\cite{Bachir1}. 
 
The model Eq.~(\ref{eq:Fphase}) is only valid in the $\tau$ decay region. 
Thus, in Eq.~(\ref{eq:Dispfv}) the phase is
taken as
\begin{equation}
\phi_+(s) = \left \{
\begin{array}{l}
{\rm{tan}}^{-1} \left[ \frac{{\rm{Im}}~\tilde f_+(s) }{{\rm{Re}}~\tilde f_+(s)}\right]~, \,\,\, s \leq  s_{cut}  \nonumber \\
\pi \pm \pi, \,\,\, s \geq  s_{cut}
\label{eq:delta}
\end{array} 
\right .
\end{equation}
with $s_{cut}$ of the order of $m_\tau^2$. 
For $s \geq  s_{cut}$, we use the asymptotic value of $\phi_+$ with a large error band. The interest
of using a three time subtracted dispersion relation is that the impact of 
our ignorance of the phase at relatively high e-nergy turns out to be very small.
Using such a model, two sum-rules dictated by the asymptotic behaviour of $\bar f_+(s)$ have to be fulfilled
\begin{equation}
\lambda_+' = -\frac{m_\pi^2}{\pi}  \int_{s_{K\pi}}^{\infty} ds' \frac{\phi_+(s')}{s'^2}~,
\label{eq:sumrule1}
\end{equation}  
\begin{equation}
\lambda_+'' - \lambda_+'^2 = \frac{2 m_\pi^4}{\pi}  \int_{s_{K\pi}}^{\infty} ds' \frac{\phi_+(s')}{s'^3}~. 
\label{eq:sumrule2}
\end{equation} 
If $\phi_+(s)$ was exactly known, these two sum-rules would allow for a determination of the two subtraction constants $\lambda_+'$ and 
$\lambda_+''$. In our fits, these relations yield additional constraints on the parameters 
especially the second one where the influence of the high-energy region is suppressed. 
\subsection{Scalar form factor}
Analogously to our discussion for the vector form factor, we write a dispersion relation for ln$\bar f_0(s)$ with three subtractions.  
Motivated by the existence of the CT theorem, one subtraction is performed at the 
CT point where we would like to determine the form factor and the other two at $s=0$. 
This leads to the following dispersive representation for $\bar f_0(s)$
\begin{eqnarray}
\label{eq:Dispfs}
\bar f_0(s) &\!\! = \!\! & \exp \left[ \frac{s}{\Delta_{K\pi}} \left( \mathrm{ln}C+
(s-\Delta_{K\pi}) \right. \right. \\
& & \times \left( \frac{\mathrm{ln}C}{\Delta_{K\pi}} - \frac{\lambda_0'}{m_\pi^2} \right)+ 
\frac{\Delta_{K\pi}~s~(s-\Delta_{K\pi})}{\pi}   \nonumber \\
& & \left. \left. \times \int_{s_{K\pi}}^{\infty}
\frac{ds'}{s'^2}
\frac{\phi_0(s')}
{(s'-\Delta_{K\pi})(s'-s-i\epsilon)}\right) \right]~. \nonumber
\end{eqnarray}
The two subtraction constants ln$C$ = ln$\bar f_0(\Delta_{K\pi})$, see Eq.~(\ref{eq:CTrel}), 
and $\lambda_0'$, the slope of the form factor, see Eq.~(\ref{eq:Taylor})
(the third one is fixed since $\bar f_0(0)=1$), are determined from a fit to the data. $\phi_0(s)$ re-presents 
the phase of the form factor. It can be identified 
in the elastic region with the $S$-wave $I=1/2$ K$\pi$ scattering 
phase~\cite{Watson}. The latter has been extracted from the data in Ref.~\cite{Bachir2} and will be used as input in 
the dispersive parametrization, Eq.~(\ref{eq:Dispfs}). 
In the inelastic region or high ener-gy region (for $s \geq s_{in} \equiv 2.77$ GeV$^2$) 
where the phase is unknown a large band of $2\pi$ is considered 
for the phase ($\phi_{0,\mathrm{as}}(s) \equiv (\pi \pm \pi)~\theta(s-s_{in})$). 
Note that compared to the dispersive parametrization proposed in Refs.~\cite{BOPS06,BOPS09}, one more subtraction is needed 
since the $\tau \rightarrow K \pi \nu_\tau$ decays take place at much higher energy than the $K_{\ell 3}$ decays. This allows to have the 
theoretical uncertainties from the high energy phase under control, the phase being suppressed by $1/s'^3$ in the dispersive integral, 
Eq.~(\ref{eq:Dispfs}). 
In order for the form factor to have the correct asymptotic behaviour, the following sum-rules should be fulfilled
\begin{equation}
{\mathrm{ln}C} = \frac{\Delta_{K\pi}}{\pi} 
\int_{s_{K\pi}}^{\infty} \frac{ds'}{s'} \frac{\phi_0(s')}{(s'-\Delta_{K\pi})}~,
\label{eq:SumRulefs1}
\end{equation}
\begin{equation}
\frac{\mathrm{ln}C}{\Delta_{K\pi}}- \frac{\lambda_0'}{m_\pi^2}= \frac{\Delta_{K\pi}}{\pi} 
\int_{s_{K\pi}}^{\infty} \frac{ds'}{s'^2} \frac{\phi_0(s')}{(s'-\Delta_{K\pi})}~.
\label{eq:SumRulefs2}
\end{equation}
While the constraint given by the sum-rule Eq.~(\ref{eq:SumRulefs1}) is easy to fulfill due to the large band taken for $\phi_{0,\mathrm{as}}(s)$, 
this is not the case anymore for the constraint given by Eq.~(\ref{eq:SumRulefs2}) 
which plays an important role in the determination of the two unknowns ln$C$ and $\lambda_0'$ 
from the fit to the data. 
\section{FITS TO THE $\tau \rightarrow K\pi \nu_\tau$ AND $K_{\ell3}$ DATA}
\subsection{Presentation}
The $\tau \rightarrow K \pi \nu_\tau$ decay spectrum has been
measured by Belle~\cite{Belle} and BaBar~\cite{BaBar}. The Belle data\footnote{We would like 
to acknowledge D. Epifanov for providing us with the Belle spectrum.} are shown in Fig.~1. 
The number of events in a given bin i is given by~\cite{Jamin}
\begin{equation}
N(i)=N_{tot}~b_{w}~\frac{1}{\Gamma_{K\pi}} \frac{d\Gamma_{K\pi}}{d\sqrt{s}}(s_i)~,
\label{eq:Nbin}
\end{equation}
with $N_{tot}$, the total number of events, $b_{w}$ the bin width and $\Gamma_{K\pi}$ the decay width given by Eq.~(\ref{eq:M}). 
An important remark here is that in Eq.~(\ref{eq:Nbin})  the normalization, see Eq.~(\ref{eq:M}), cancels by taking the ratio $1/\Gamma_{K\pi}~d\Gamma_{K\pi}/d\sqrt{s}$.
Thus, in order to fit the data one does not need to know \Vus. 
We use for the two form factors the dispersive parametrizations 
of Eqs.~(\ref{eq:Dispfv})~and~(\ref{eq:Dispfs}) to fit the spectrum up to $s_\mathrm{fit} \sim (1.5$ GeV$)^2$, see Fig.~1. 
Indeed, above this energy theoretical as well as experimental uncertainties start to become important. 
Nine para-meters for the form factors are determined. Two 
for $\bar f_0 (s)$, lnC=ln$\left(\bar f_0(\Delta_{K\pi})\right)$ and $\lambda_0'$ and seven 
for $\bar f_+ (s)$, $\lambda_+'$, $\lambda_+''$, the mass and width of the $K^*$ and $K^{*'}$ 
resonances $m_{K*}$, $\Gamma_{K*}$, 
$m_{K*'}$, $\Gamma_{K*'}$ and $\beta$, the mixing parameter. 
We add in the fits the cons-traints given by the sum-rules
Eqs.~(\ref{eq:sumrule1},\ref{eq:sumrule2},\ref{eq:SumRulefs1},\ref{eq:SumRulefs2}).
Once the form factors are determined, one can compute the 
phase space integrals and extract $f_+(0)$ \Vus~from the decay width measurement, Eq.~(\ref{eq:M}).  
\begin{figure}[t]
\hspace{-1.05cm}
\includegraphics[width=0.55\textwidth]{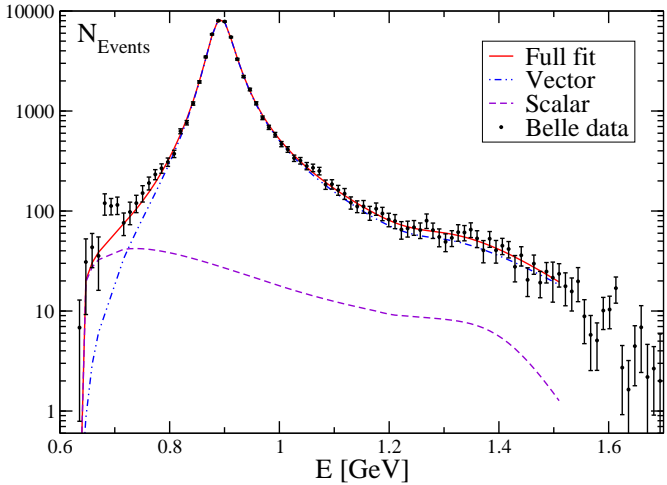}
\vspace{-1.4cm}
\caption{\it Fit result for the spectrum of \mbox{$\tau \rightarrow K \pi \nu_\tau$}. 
The data in black are from Belle Collaboration~\cite{Belle}. 
The dashed violet line represents the scalar form factor contribution fixed from the $K_{\mu 3}$ results, see text. 
The dot-dashed blue line is the vector form factor contribution and the solid red line gives the full result.}
\label{fig:Taus}
\vspace{-0.5cm}
\end{figure}

The introduction of the dispersive parametrizations Eqs.~(\ref{eq:Dispfv},\ref{eq:Dispfs}) which are valid 
in the full energy range allows 
to combine the $\tau \rightarrow K \pi \nu_\tau$ decay analysis with 
the $K_{\ell 3 }$ one to further constrain the K$\pi$ form factors. Note that for the moment 
we cannot take into account the results on the vector form factor parameters coming from 
$K_{\mu 3}$ decays because we do not have the correlations between ln$C$ and $\lambda_+'$ and $\lambda_+''$. 
In the future, a combined fit of the  
$\tau$ and $K_{\ell 3}$ decays should be performed using the same dispersive parametrization in order to take into account 
all the available data and the correlations between the two form factors.   
\subsection{Discussion}
Since our fits are still preliminary we refrain from quoting final
results. Instead, we concentrate in a discussion of the prospects of
our \mbox{analysis~\cite{BDP}}. To do so, we show in Fig.~1 
the contribution of the scalar form factor to the $\tau \rightarrow K \pi \nu_\tau$ decay spectrum, Eq.~(\ref{eq:Nbin}), 
where the  value for ln$C$ has been taken from the $K_{\mu 3}$ analyses~\cite{Flavia}, 
\mbox{$\mathrm{ln}C= 0.2004 \pm 0.0091$}, and $\lambda_0'$ has been determined from the sum-rule Eq.~(\ref{eq:SumRulefs2}),
\mbox{$\lambda_0'= 13.71 \cdot 10^{-3}$}. The vector form factor has been fitted to the 
data with the two scalar form factor parameters fixed to the later values. 
Its contribution is also shown in Fig.~1 together with the total contribution to the decay spectrum. 
As can be seen, some information on $\bar f_0(s)$ can be obtained 
from $\tau \rightarrow K \pi \nu_\tau$ close to threshold ($s_{K\pi}$).  
But at present the Belle data alone are not precise enough 
to really be able to give strong constraints on $\bar f_0(s)$. 
A measurement of the forward-backward asymmetry would be very useful to di-sentangle the scalar and vector 
form factors~\cite{Truong}.   
As it has been already shown in Refs.~\cite{Belle,BaBar,Jamin,Bachir1,Diogo1}, the $\tau \rightarrow K \pi \nu_\tau$ decay spectrum 
measurement gives interesting constraints on $\bar f_{+}(s)$ and in particular on the mass and width of 
$K^*(892)$. Note that in the Belle data, Fig.~1, there is a bump close to threshold given by three points, bins 6, 7 and 8 
which cannot be accommodated by the form factor parametrizations and which does not seem to be present in the BaBar data~\cite{BaBar}. 
Awaiting the more precise measurements of the \mbox{$\tau \rightarrow K \pi \nu_\tau$} decays that are underway, an interesting 
possibility offered by the dispersive parametrization is to combine the $\tau$ decay analyses with the $K_{\ell 3}$ decays~\cite{Diogo2} and test 
the consistency of the determinations of the form factor parameters. 
As shown in Ref.~\cite{Diogo2}, it allows for a very precise determination of $\lambda_+'$ and $\lambda_+''$ since the correlations 
of these two parameters are of opposite sign in the two analyses. As for $\bar f_0(s)$, the combination allows for determining in addition to 
ln$C$, $\lambda_0'$ directly from the data. 
Last but not least, this analysis offers a direct extraction of \Vus~from $\tau \rightarrow K \pi \nu_\tau$ decays 
and an interesting consistency-test of the determination of \Vus~from $\tau$ decays 
by comparing its value to the one coming from inclusive hadronic $\tau$ decays.  
\section{CONCLUSION}
With the new measurements of $\tau \rightarrow K \pi \nu_\tau$ at the B factories~\cite{Belle,BaBar} and the forthcoming ones~\cite{BESSIII}, 
a precise extraction of the K$\pi$ form factors becomes possible. To this end, we have built a \mbox{physically} well-motivated 
dispersive representation for the form factors. One interesting feature of this parametrization is that it allows to combine the 
$K_{\ell 3}$ and $\tau \rightarrow K \pi \nu_\tau$ analyses in order to increase the precision in the determination 
of the form factor parameters.  
This allows for stringent tests of the Standard Model and in particular for an extraction of \Vus~directly from 
$\tau \rightarrow K \pi \nu_\tau$ decays.   

\section*{ACKNOWLEDGMENTS}
We would like to thank the organizers for this very pleasant 
conference. We are grateful to 
B. Moussallam, J. Portol\'es and B. Shwartz for interesting discussions and M. Jung for a careful reading of the manuscript. 
This work has been supported in part by MEC, Spain (grants FPA2007-60323 and Consolider-Ingenio 2010 CSD2007-00042, CPAN), 
by MICINN, Spain (grant CICYT-FEDER-FPA2008-01430), by the EU Contracts MRTN-CT-2006-035482 (FLAVIAnet) 
and Hadron-Physics2 (grant n. 227431) 
and by Generalitat Valenciana (PRO-METEO/2008/069).

\end{document}